\documentclass[conference,compsoc]{IEEEtran}
\usepackage[utf8]{inputenc}
\usepackage[T1]{fontenc}
\usepackage{amsmath,amssymb}
\usepackage{booktabs}
\usepackage{xcolor}
\usepackage[hidelinks]{hyperref}

\IEEEoverridecommandlockouts
\begin{document}

\title{Building Weighted PubMed Citation Graphs at Scale: A Pipeline and an Audit of Edge-Similarity Features}

\author{\IEEEauthorblockN{Hossein Mohasel Arjomandi\thanks{This work is not related to the author's current employment at Microsoft. The views expressed are the author's own. This work has been submitted to the IEEE for possible publication. Copyright may be transferred without notice, after which this version may no longer be accessible.}}
\IEEEauthorblockA{Microsoft, Redmond, WA, USA\\
hosseinm@microsoft.com}
}

\maketitle

\begin{abstract}
Studies of scientific literature often combine citation links with article metadata, but the data preparation needed to do so at the scale of millions of papers is laborious and error-prone.
This paper describes a pipeline that turns PubMed metadata and a citation network into a graph whose edges carry seven similarity features: bibliographic coupling and co-citation (each as a count and as a Jaccard coefficient), publication-year proximity, MeSH-term similarity, and similarity of MedCPT title--abstract embeddings.
Applied to a citation network derived from the cell biology research literature, the pipeline yields 7,445,898 nodes and 57,512,086 edges with complete year, MeSH, title, and abstract metadata.
The paper reports how nodes are matched across sources, how duplicate identifiers are resolved, and which intermediate tables grew to hundreds of gigabytes.
An audit of the full feature table shows that 47.2\% of the edges share neither references nor citing documents, so their topological features are zero, and that the MeSH feature is nonzero on only 0.28\% of edges.
Where both kinds of signal exist, they agree: edges with higher coupling also have more similar abstracts and closer publication years.
In a pilot, replacing the median by a mean or best-match aggregation makes the MeSH feature nonzero on nearly all article pairs, and near-identical texts point to possible duplicate records that identifier checks miss.
The parsing, graph-construction, and feature code are public; the paper states which data are released and which are not.
A companion paper uses the resulting weighted graph for clustering.
\end{abstract}


\section{Introduction}
Literature maps, topic clustering, and funding-portfolio analysis often rest on a citation graph.
Citation links alone are noisy, and biomedical articles carry additional metadata: MeSH indexing terms, publication year, and abstracts.
Combining the two requires joining identifiers across sources, handling duplicates, and computing pairwise features for tens of millions of edges.
These steps are easy to get wrong, and several of them dominate storage and run time.
The intended users are researchers who build weighted citation graphs from PubMed for science mapping, such as clustering, and who need to know what each step costs and where it can fail.

This paper describes one such pipeline and the data it produces.
The contributions are:
\begin{itemize}
\item a stage-by-stage description of the pipeline that builds a PubMed citation graph with complete metadata, with record counts at each stage (Section~\ref{sec:pipeline});
\item definitions of seven edge-similarity features, with the reason for each design choice (Section~\ref{sec:features});
\item an audit of the full feature table: coverage, ranges, distributions, correlations, agreement between topological and semantic features, a pilot test of MeSH aggregations, and near-duplicate detection from text similarity (Sections~\ref{sec:audit}--\ref{sec:lessons});
\item engineering observations on duplicate identifiers, feature coverage, and intermediate storage (Section~\ref{sec:lessons});
\item a statement of which data and code are released (Section~\ref{sec:availability}).
\end{itemize}
The graph was built for a clustering study, which is reported separately~\cite{Arjomandi_B}; this paper does not report clustering results.
The work derives from the author's M.S. thesis~\cite{Arjomandi_2024}.

\section{Related Work}
\label{sec:related}
Relatedness between publications is commonly measured by direct citation, bibliographic coupling~\cite{Kessler_1963}, co-citation~\cite{Small_1973}, or text similarity, and these measures have been compared for clustering publications~\cite{Waltman_2020,Boyack_2011}.
Ahlgren et al.~\cite{Ahlgren_2020} enhance direct citations with bibliographic coupling, co-citation, extended direct citations, and the BM25 text measure, and evaluate the resulting clusterings against MeSH; Boyack and Klavans~\cite{Boyack_2020} compare direct-citation, textual, and hybrid relatedness on nine million documents.
Those studies ask which measure to use. This paper describes the data engineering that produces such features for a graph of several million nodes, and the resulting data.
Existing resources supply the entities and links that such a pipeline starts from: the NIH Open Citation Collection for citations among PubMed articles~\cite{Hutchins_2019}, COCI for open DOI-to-DOI citations~\cite{Heibi_2019}, S2ORC for papers with full text and citation links~\cite{Lo_2020}, and the PubMed Knowledge Graph for PubMed abstracts linked to authors, affiliations, and funding~\cite{Xu_2020}.
MedCPT~\cite{Jin_2023} provides pre-computed embeddings of PubMed titles and abstracts~\cite{MedCPT_embeddings}, which serve here as one feature.

\section{Data Sources}
\label{sec:data}
\emph{Citation network.}
The starting point is a curated citation network (CEN) with about 14 million nodes and 92 million edges~\cite{Arjomandi_2024}, derived from the cell biology literature and centered on extracellular vesicles (exosomes); its edge list, with integer node IDs, is public~\cite{Park_2023_data}. Other analyses of such networks include overlapping-community detection~\cite{Jakatdar_2022} and a core--periphery study~\cite{Wedell_2022}.
Nodes carry integer identifiers and DOIs.

\emph{PubMed metadata.}
Metadata come from the PubMed annual baseline, which is a complete snapshot of PubMed citation records in XML~\cite{Arjomandi_2024}.
A public table of parsed 2024 metadata, restricted to records that have a PMID, a DOI, a title, and an abstract, produced with the same parsing code, is part of the dataset ``Parsed Open Citations and PubMed Data''~\cite{Arjomandi_2024_data} (DOI~\texttt{10.13012/B2IDB-5216575\_V1}, CC~BY~4.0).
The same dataset also provides an integer-ID edge list of the January 2024 Open Citations release with 121,052,490 nodes and 1,962,840,983 edges; this edge list is not used for the graph described here.

\emph{Embeddings.}
The title--abstract embeddings (768 dimensions) released with MedCPT~\cite{Jin_2023,MedCPT_embeddings} are used.

\section{Pipeline}
\label{sec:pipeline}
Table~\ref{tab:stages} summarizes the stages. Counts in stages 2 and 6--7 (number of nodes with lists) are taken from the author's SQL and processing logs; the edge and row counts in stages 4--7 were recomputed from the final feature table (Section~\ref{sec:audit}) and match the logs.

\begin{table}[t]
\centering
\caption{Pipeline stages and record counts.}
\label{tab:stages}
\small
\begin{tabular}{p{0.2cm}p{4.2cm}p{2.9cm}}
\toprule
\# & Stage & Output / count\\
\midrule
1 & Parse PubMed XML to parquet (DOIs lower-cased); keep one record per DOI; keep latest revision per PMID; drop records without DOI & metadata table\\
2 & Join CEN nodes to PubMed by DOI; keep records with year, MeSH, title, and abstract & 8,599,582 node--PMID rows (before PMID de-duplication)\\
3 & Resolve repeated PMIDs by keeping the node of highest degree & one node per PMID\\
4 & Keep edges whose endpoints both survive; keep nodes with at least one edge & 57,512,086 edges; 7,445,898 nodes (graph called CENM below)\\
5 & Year and MeSH features per edge (Spark) & 57,512,086 rows\\
6 & Bibliographic coupling from out-neighbor lists (6,415,071 nodes have one) & 53,200,439 edges\\
7 & Co-citation from in-neighbor lists (4,120,853 nodes have one) & 44,509,707 edges\\
8 & Join MedCPT embeddings by PMID; remove duplicate rows & 7,445,837 nodes\\
9 & Scale features and combine with user weights & weighted edge list\\
\bottomrule
\end{tabular}
\end{table}

\emph{Stage 1.}
The extraction is a PySpark job that writes distributed parquet files; the repository README states a run time of about one hour and suggests 70 cores for the server it was developed on~\cite{pubmed_etl_repo}.
The parser lower-cases DOIs. The job then keeps one record per DOI (Spark \texttt{dropDuplicates}, which does not control which record is kept), keeps for each PMID the record with the latest revision date, and drops records with an empty DOI. Because the DOI step runs first, it can in principle drop the latest revision of a PMID (see Limitations); running the PMID step first avoids this.

\emph{Stages 2--4.}
Matching is an exact join of the lower-cased PubMed DOIs with the DOIs of the CEN node list; CEN DOIs that differ only in case would be lost (see Limitations).
Because one DOI can correspond to several node identifiers and one PMID can appear for several nodes, the node with the highest degree in the induced subgraph is kept for each PMID, which also keeps the graph free of duplicate articles.
Only edges with both endpoints retained are kept, and the graph, called CENM (CEN with metadata), is not modified afterwards.

\emph{Stages 6--7.}
Features that depend on neighbor sets need, for each edge, the neighbor lists of both endpoints.
Of the 7,445,898 nodes, 6,415,071 cite at least one other node of the graph and 4,120,853 are cited by at least one. An edge gets a row only if both endpoints have the needed list, which explains why fewer than 57.5 million edges appear in stages 6 and 7. A missing row means that one endpoint has an empty reference (or citing) set, so the true overlap is zero; setting these features to zero is therefore exact, not an imputation (Section~\ref{sec:missing}).

\emph{Stage 8.}
Duplicate embedding rows were reduced to one row per node by random selection, leaving 7,445,837 embeddings for 7,445,898 nodes, so 61 nodes have no embedding. The text-similarity job assigns the constant $0.5$ to any edge for which either endpoint has no usable embedding (Section~\ref{sec:missing}).

\section{Edge Features}
\label{sec:features}
Let $(a,b)$ be a citation edge, treated as undirected, $R_x$ the set of references of $x$ within the graph, and $C_x$ the set of documents citing $x$.
The topological features are
\begin{align}
f_1 &= |R_a \cap R_b|, & f_2 &= \frac{|R_a \cap R_b|}{|R_a \cup R_b|},\\
f_3 &= |C_a \cap C_b|, & f_4 &= \frac{|C_a \cap C_b|}{|C_a \cup C_b|},
\end{align}
the bibliographic-coupling count and Jaccard coefficient, and the co-citation count and Jaccard coefficient.
Three semantic features follow.
Year similarity $f_5$ equals $1$ if the publication years differ by at most five, $0.5$ if they differ by more than five and at most ten, and $0.25$ otherwise. The thresholds follow the thesis, which motivates them by the concentration of citations on recent work~\cite{Price_1965,Silva_2020}; they were not tuned.
Text similarity $f_6$ is the cosine similarity of the MedCPT embeddings of the two title--abstract texts. Cosine similarity can be negative in general, and the code does not clip it; on CENM the observed values lie in $[0.498, 1]$ (Section~\ref{sec:audit}).
MeSH similarity $f_7$ uses the MeSH tree: each term maps to one tree number, taken from a lookup of 29,944 descriptors built from NLM's MeSH pages~\cite{Arjomandi_2024} (descriptors with several tree numbers keep only the stored one), and the similarity of two terms is the number of shared leading path components, counting the leading letter as one component. For example, \texttt{D12.776.828.300} and \texttt{D12.776.4311.09.1206} share three. $f_7$ is the median of this value over all term pairs of the two articles; terms missing from the lookup are skipped, and $f_7=0$ if no pair remains. Two terms from different top-level MeSH categories share nothing, so $f_7=0$ whenever more than half of the term pairs cross categories; this is one reason $f_7$ is often zero (Section~\ref{sec:audit}).

\emph{Toy example (hypothetical terms, not real MeSH data).}
Let article $a$ carry the terms \texttt{A01.100.200} and \texttt{A01.100.300}, and article $b$ the single term \texttt{A01.100.200.50}.
Splitting after the leading letter gives the components $(\mathrm{A},01,100,200)$, $(\mathrm{A},01,100,300)$, and $(\mathrm{A},01,100,200,50)$.
The first pair shares four leading components and the second shares three, so $f_7$ is the median of $\{4,3\}$, i.e.\ $3.5$, before the network-wide scaling.

Features $f_1$--$f_4$ and $f_7$ are min--max scaled to $[0,1]$ over the whole network (Spark \texttt{MinMaxScaler}) after missing values are set to zero. Their minimum is $0$ on CENM, so the scaling equals division by the maximum, which is the max-normalization described in the thesis.
$f_5$ and $f_6$ already lie in $[0,1]$ and are used as computed. The features therefore differ in scale: $f_5$ and $f_6$ are large on every edge, while the scaled counts are small on most edges (Section~\ref{sec:audit}), so with comparable weights $f_6$ contributes a nearly constant term. Rescaling $f_6$ over its observed range before weighting is recommended.
Min--max scaling was used for simplicity. For heavy-tailed counts it compresses most values toward zero, and a rank or logarithmic transform would be an alternative that was not tested.
With weights $w_i\ge0$, $\sum_i w_i=1$, the edge weight is $s(a,b)=\sum_{i=1}^{7} w_i f_i(a,b)$.

\subsection{Edge record}
Table~\ref{tab:schema} lists the columns of the per-edge feature table as named in the processing scripts, before the weights are applied.

\begin{table}[t]
\centering
\caption{Columns of the per-edge feature table (from the processing scripts). Scaled: min--max scaled over the network.}
\label{tab:schema}
\small
\begin{tabular}{@{}lcl@{}}
\toprule
Column & Feature & Value\\
\midrule
\texttt{first}, \texttt{second} & -- & node IDs\\
\texttt{bib\_frequency} & $f_1$ & scaled count\\
\texttt{bib\_jaccard} & $f_2$ & scaled, $[0,1]$\\
\texttt{cocitation\_frequency} & $f_3$ & scaled count\\
\texttt{cocitation\_jaccard} & $f_4$ & scaled, $[0,1]$\\
\texttt{year\_similarity} & $f_5$ & $\{1,0.5,0.25\}$\\
\texttt{cosine\_similarity} & $f_6$ & as computed\\
\texttt{mesh\_median} & $f_7$ & scaled median\\
\texttt{weight} & $s$ & weighted sum\\
\bottomrule
\end{tabular}
\end{table}

\subsection{Computation}
\label{sec:comp}
The two set-overlap features are computed per edge from the neighbor lists of the two endpoints.
The lists are stored as comma-separated identifier strings, split and sorted, and intersected with a two-pointer merge, so one edge costs time linear in the two list lengths.
The Jaccard coefficient is the intersection size divided by $|A|+|B|-|A\cap B|$.
Bibliographic coupling ($f_1$, $f_2$), year, and MeSH similarity are computed with Spark jobs over the edge table.
Co-citation ($f_3$, $f_4$) is computed by a separate Python script that reads the large in-neighbor table in chunks of 10,000 rows, processes the chunks with a pool of 40 worker processes, and inserts the results into the database.
The text similarity $f_6$ is the cosine of two stored vectors, computed in a similar chunked script.

\subsection{Missing-value rules}
\label{sec:missing}
Three rules decide the value of a feature when its inputs are absent, and they affect the weights.
(i) If an endpoint has no neighbor list in the graph, the edge has no row in the corresponding feature table, so its value is missing in the merged edge table and is replaced by zero in the normalization step. This zero equals the true value, because the endpoint's reference (or citing) set is empty.
(ii) If either endpoint has no embedding, or a stored value with at most 100 entries, the text similarity is set to the constant $0.5$. The length test guards against empty or malformed stored values; valid embeddings have 768 entries.
(iii) Year similarity is always defined, because the graph keeps only articles with a year. MeSH similarity is $0$ when none of the term pairs can be looked up in the tree.
Rule (i) uses zero because zero is the exact value. For rule (ii), $0.5$ was an engineering default for a handful of nodes. It is not a neutral value: it lies at the bottom of the observed cosine range, so it acts as low similarity. Exactly 61 nodes have $f_6=0.5$ on every incident edge, consistent with the 61 nodes without an embedding, and they touch 198 edges, which are all edges with $f_6=0.5$. Rule (ii) thus affects 198 edges (0.0003\%), so its effect on the weights is negligible.

\emph{Checks used during the build.}
The SQL log records these checks: row, node, and PMID counts after the join to PubMed; equal numbers of distinct PMIDs and nodes after de-duplication; a query for repeated (first, second) pairs in the co-citation table; and, for the parsing, equal numbers of parquet and XML files~\cite{pubmed_etl_repo}.

\section{Audit of the Feature Table}
\label{sec:audit}
Unlike Sections~\ref{sec:pipeline}--\ref{sec:features}, this audit was not part of the thesis~\cite{Arjomandi_2024}. The final per-edge table (57,512,086 rows, raw feature values before scaling) was read, and coverage, ranges, distributions, rank correlations, and an agreement check were computed. The audit and MeSH pilot (Section~\ref{sec:mesh}) scripts, with configuration, seed, results, and the CSV summaries behind Tables~\ref{tab:dist} and~\ref{tab:agree}, are public~\cite{audit_repo}.

\subsection{Coverage and distributions}
Table~\ref{tab:coverage} gives the number of edges with a coupling or co-citation row and the number with a nonzero topological weight. All three counts agree with the logs and with the thesis.

\begin{table}[t]
\centering
\caption{Edge coverage of the features (57,512,086 edges), recomputed from the feature table. The rows and the nonzero count match the processing logs and~\cite{Arjomandi_2024}, Table~4.3.}
\label{tab:coverage}
\small
\begin{tabular}{p{3.9cm}rr}
\toprule
Feature rows & Edges & Share (\%)\\
\midrule
Year and MeSH (all edges) & 57,512,086 & 100.0\\
Bibliographic-coupling row & 53,200,439 & 92.5\\
Co-citation row & 44,509,707 & 77.4\\
Nonzero topological weight (equal weights) & 30,382,764 & 52.8\\
\bottomrule
\end{tabular}
\end{table}

Table~\ref{tab:dist} summarizes each feature. The counts are heavy-tailed: the largest co-citation count is 5,811 while the median nonzero count is 2, so after min--max scaling 99.2\% of the nonzero co-citation counts fall below $0.01$. The Jaccard versions are less compressed. Year similarity is $1$ on 45.0\% of the edges, $0.5$ on 28.6\%, and $0.25$ on 26.3\%. Text similarity has a median of 0.790 and an interquartile range of $[0.751, 0.827]$; only 2 edges are below $0.5$, and 61 are at least $0.99$ (Section~\ref{sec:lessons}).
\subsection{MeSH aggregation}
\label{sec:mesh}
MeSH similarity is nonzero on only 159,124 edges (0.28\%), so with the median aggregation $f_7$ is close to absent.
The feature code computed both the median and the mean of the term-pair values and stored both in the build database. The mean was dropped before scaling, and the copy of the table available to the author holds only the median. Recomputing either needs the node-to-PMID mapping, which is in the build database. The aggregation was therefore tested on the thesis pilot data: the MeSH terms of 1,591 articles of one CEN cluster, retrieved from PubMed in 2023, over all 1,264,845 article pairs (not only citing pairs). None of the 18,576 term occurrences is missing from the lookup, while 77.8\% of term pairs cross top-level categories on average. The median is nonzero for 0.14\% of the pairs.
This is not directly comparable with the 0.28\% on CENM, because CENM uses citing pairs across the whole graph, which are likely more similar than arbitrary pairs within a cluster. The two values are nevertheless of the same order, and since no lookup fails, cross-category pairs, not missing lookups, explain the sparsity. The mean of the term-pair values and a symmetric best-match average (each term matched to its most similar term in the other article) are nonzero for 99.998\% of the pairs, with interquartile ranges $[0.37, 0.57]$ and $[2.11, 3.19]$.
Replacing the median by the mean, which the feature code already computes, or by best match is recommended. Otherwise $f_7$ should be dropped.

\begin{table}[t]
\centering
\caption{Feature distributions on CENM (raw values). Median and P99 are over edges with a nonzero value. Last column: share of nonzero values below $0.01$ after min--max scaling.}
\label{tab:dist}
\footnotesize
\setlength{\tabcolsep}{3pt}
\begin{tabular}{@{}lrrrrr@{}}
\toprule
Feature & Nonzero \% & Median & P99 & Max & ${<}0.01$ \%\\
\midrule
$f_1$ coupl.\ count & 43.6 & 2 & 16 & 182 & 41.1\\
$f_2$ coupl.\ Jaccard & 43.6 & 0.028 & 0.182 & 0.983 & 11.2\\
$f_3$ cocit.\ count & 26.1 & 2 & 51 & 5,811 & 99.2\\
$f_4$ cocit.\ Jaccard & 26.1 & 0.011 & 0.167 & 0.812 & 41.7\\
$f_5$ year & 100.0 & 0.5 & 1 & 1 & --\\
$f_6$ text (cosine) & 100.0 & 0.790 & 0.906 & 1.000 & --\\
$f_7$ MeSH & 0.28 & 1 & 2 & 6 & 0.0\\
\bottomrule
\end{tabular}
\end{table}

\subsection{Correlations}
On a random sample of 2,000,000 edges (fixed seed), the Spearman correlation between count and Jaccard is 0.97 for coupling and 0.98 for co-citation, so each pair carries nearly the same ranking. Users should therefore weight only one feature of each pair; giving weight to both effectively doubles the weight of coupling or co-citation. Coupling and co-citation correlate at about 0.31--0.32, text similarity correlates with the four topological features at 0.29--0.35, and year similarity at 0.19--0.26. MeSH similarity correlates with every other feature below 0.03, as expected from its sparsity.

\subsection{Agreement between topological and semantic features}
If the features measure relatedness, edges with stronger topological overlap should also have more similar texts. Table~\ref{tab:agree} groups edges by their coupling Jaccard coefficient. Mean text similarity rises from 0.768 for edges with no topological overlap to 0.830 in the top quintile, and the share of edges within five years rises from 34.6\% to 70.9\%. The difference in $f_6$ between the first and last groups is 0.062, about 81\% of the interquartile range of $f_6$ and a large effect (Cohen's $d=1.20$). Part of the year trend is expected mechanically, because articles that cite the same references tend to be published close in time. The check therefore shows that the features are consistent with one another, not that they match an external standard.

\begin{table}[t]
\centering
\caption{Semantic features by topological overlap. Quintiles are over the 25,092,927 edges with nonzero coupling Jaccard; they differ in size because of tied values. Edges with co-citation but no coupling are in neither group.}
\label{tab:agree}
\footnotesize
\begin{tabular}{@{}lrrr@{}}
\toprule
Group & Edges & Mean $f_6$ & $|\Delta Y|\le5$ (\%)\\
\midrule
No topological overlap & 27,129,322 & 0.768 & 34.6\\
Coupling Jaccard Q1 & 4,974,430 & 0.794 & 50.0\\
Q2 & 5,056,617 & 0.798 & 50.8\\
Q3 & 4,925,344 & 0.804 & 53.3\\
Q4 & 5,117,685 & 0.813 & 58.7\\
Q5 (highest) & 5,018,851 & 0.830 & 70.9\\
\bottomrule
\end{tabular}
\end{table}

\section{Observations from Building the Graph}
\label{sec:lessons}
\emph{Intermediate tables dominate storage.}
Materializing, for every edge, the in-neighbor lists of both endpoints produced a table of about 240~GB; the corresponding out-neighbor table was also very large and was deleted after use, and a table duplicating embeddings per edge reached about 450~GB and was also deleted.
Computing the co-citation feature required an indexed primary key on the large table and chunked processing. Computing set intersections from node-level lookup tables, rather than per-edge copies of neighbor lists, could avoid most of this storage. This alternative was not tested.

\emph{Identifiers are not unique.}
Both DOIs and PMIDs repeated, and a naive join silently duplicates articles. A cheap guard is to check, after each join, that the number of distinct node identifiers equals the number of distinct PMIDs. It does not catch the same work under two PMIDs: 61 citation edges link texts with cosine similarity of at least 0.99, 49 of them within five years. The one edge with similarity $1$ joins two articles of the same year that share 48 references (coupling Jaccard 0.89). None of these edges touches a node without an embedding, so the count of 61 is unrelated to the 61 missing embeddings. These pairs may be republications, corrigenda, or duplicate records, but this cannot be settled without the texts.

\emph{Many edges carry no topological signal.}
47.2\% of the edges have zero topological weight (Table~\ref{tab:coverage}). These edges are direct citations between articles that share no references and no citing documents. A weighting that uses only topological features gives them weight zero, which removes them from a weighted objective. In the companion study~\cite{Arjomandi_B}, these edges hold together most of the clustering coverage.

\emph{Audit the features, not only the pipeline.}
The sparsity of the median-based MeSH feature (Table~\ref{tab:dist}) was invisible during the build, where every stage produced the expected number of rows. A per-feature distribution check before weighting would have caught it.

\section{Discussion}
A user who wants a weighted PubMed graph today has three options. A bibliographic database such as OpenAlex~\cite{Priem_2022} supplies citations, identifiers, MeSH terms, abstracts, and a list of related works per paper but, to the author's knowledge, no per-edge similarity scores, so the overlap computations and scaling choices fall to the user. PubMed's related-articles measure (PMRA), evaluated for clustering by Boyack et al.~\cite{Boyack_2011}, gives one relatedness score per pair, not separable features. Publication-level classification systems such as that of Waltman and van Eck~\cite{Waltman_2012} produce clusters, not per-edge features.
This pipeline addresses the gap between these options. It documents each step from identifier matching to scaling (Sections~\ref{sec:pipeline} and~\ref{sec:features}), gives record counts at every stage (Table~\ref{tab:stages}) against which a rebuild can be checked, and includes an audit that shows which features carry signal. Count and Jaccard versions of a topological feature rank edges almost identically (Spearman 0.97 and 0.98), so only one of each pair needs weight. Of the edges, 47.2\% carry no topological signal, and the median MeSH feature is nonzero on 0.28\%, so a different aggregation, or dropping it, is advisable.

\section{Reproducibility Checklist}
Table~\ref{tab:repro} lists what is needed to reproduce the graph and what is currently available.

\begin{table}[h]
\centering
\caption{Reproducibility checklist.}
\label{tab:repro}
\small
\begin{tabular}{p{2.3cm}p{5.0cm}}
\toprule
Item & Status\\
\midrule
Source data & PubMed metadata: public~\cite{Arjomandi_2024_data}; CEN edge list (integer IDs): public~\cite{Park_2023_data}; node-to-article map: not released\\
PubMed parsing & public repository (\texttt{pubmed\_etl})\\
Software & Parsing: Python 3.6.8, PySpark 3.2.4, pandas 1.1.5, NumPy 1.19.5, psycopg2 2.9.8, PostgreSQL JDBC 42.5.2; clustering: \texttt{leidenalg} (docs version 0.10.3.dev0, per the thesis); others not recorded\\
Embeddings & public MedCPT release~\cite{MedCPT_embeddings}\\
CENM node/edge lists & not publicly released (derived from the research group's network data)\\
Feature table & not released; audit and MeSH pilot scripts and summary CSVs public~\cite{audit_repo}\\
Graph, feature, weighting, and clustering code & public cleaned copy~\cite{thesis_pipeline_repo} of \texttt{hm31\_graphs}; see the weight-alignment caveat in Section~\ref{sec:availability}\\
Run time & About one hour each for parsing, inserting the year and MeSH features, and building the in-neighbor edge table (SQL log); other stages not timed\\
\bottomrule
\end{tabular}
\end{table}

\section{Limitations}
The pipeline was run once on one graph (cell biology literature centered on exosomes), on one database server. Only approximate timings from the logs are reported for some stages, and there is no comparison with other pipelines; the steps are not specific to citation data, but they have not been run on another graph.
The features are checked for internal consistency (Section~\ref{sec:audit}) but not against an external relatedness benchmark.
The build database is no longer available to the author, so two effects that need the intermediate tables could not be measured: how many CEN DOIs differ from PubMed DOIs only in case, and how often the DOI de-duplication drops the latest revision of a PMID. Both affect which nodes enter the graph, not the features of the edges that do.
The MeSH feature is almost always zero with the median aggregation used here; the better aggregations were tested only on pilot data, because recomputing $f_7$ on CENM needs the node-to-PMID mapping.

\section{Data and Code Availability}
\label{sec:availability}
\emph{Released:} the parsed PubMed metadata and an Open Citations edge list (not used here)~\cite{Arjomandi_2024_data}; the PubMed parsing code~\cite{pubmed_etl_repo}; the code that builds the graph, computes the seven features, applies the weights, and runs the clustering, including the SQL log of the graph construction, as a cleaned copy~\cite{thesis_pipeline_repo} of the original repository \texttt{hm31\_graphs} (caveat: its clustering step attaches weights to edges by position, which misaligns them; see~\cite{Arjomandi_B} and the erratum in the repository README); and the audit script of Section~\ref{sec:audit} with its summary CSVs, and the MeSH pilot script of Section~\ref{sec:mesh} with its results~\cite{audit_repo}, in \texttt{pubmed-meta-etl}, the author's maintained copy of \texttt{pubmed\_etl}.
\emph{Not released:} the mapping of CEN nodes to articles (research group data), and so the CENM graph and its seven-feature table.

\section*{Acknowledgment}
The author thanks George Chacko for reading the manuscript and offering suggestions.

Use of generative AI: Claude (Anthropic) was used to draft and edit the text of all sections, to write the \LaTeX{} source including the tables, and to write the audit script of Section~\ref{sec:audit}. All numbers come from the thesis, the processing logs, the code, or the output of that script, and were checked by the author, who reviewed and validated all content and takes full responsibility for it.

\bibliographystyle{IEEEtran}
\bibliography{refs}

\end{document}